\documentstyle{article}
\newcommand{\ddouble}{{\partial^{^{\kern-6pt \leftrightarrow}}}}

\newcommand{\beq}{\begin{equation}}
\newcommand{\eqn}[1]{\label{#1}\end{equation}}
\newcommand{\ba}{\begin{array}}
\newcommand{\ea}{\end{array}}
\newcommand{\journal}[4]{{\em #1~}#2\,(19#3)\,#4;}
\newcommand{\jhep}{\journal {JHEP}}

\newcommand{\hpa}{\journal {Helv. Phys. Acta}}

\newcommand{\jmp}{\journal {J. Math. Phys.}}

\newcommand{\np}{\journal {Nucl. Phys.}}
\newcommand{\pl}{\journal {Phys. Lett.}}

\def\l{\lambda}
\def\m{\mu}
\def\n{\nu}

\def\psi{\Psi}

\def\cf{{\cal F}}

\def\inbar{\vrule height1.5ex width.4pt depth0pt}
\def\rlx{\relax\leavevmode}
\def\I{\leavevmode\hbox{\small1\kern-3.8pt\normalsize1}}
\def\openone{\leavevmode\hbox{\small1\kern-3.3pt\normalsize1}}
\def\Ione{\rlx{\rm 1\kern-2.7pt l}}
\def\Ik{\rlx{\rm I\kern-.18em k}}
\def\IC{\rlx\leavevmode
             \ifmmode\mathchoice
                    {\hbox{\kern.33em\inbar\kern-.3em{\rm C}}}
                    {\hbox{\kern.33em\inbar\kern-.3em{\rm C}}}
                    {\hbox{\kern.28em\sinbar\kern-.25em{\rm C}}}
                    {\hbox{\kern.25em\ssinbar\kern-.22em{\rm C}}}
             \else{\hbox{\kern.3em\inbar\kern-.3em{\rm C}}}\fi}
\def\IP{\rlx{\rm I\kern-.18em P}}
\def\IR{\rlx{\rm I\kern-.18em R}}
\def\IN{\rlx{\rm I\kern-.20em N}}

\def\llsymbol#1{\@llsymbol{\@nameuse{c@#1}}}
\def\@llsymbol#1{\ifcase#1\or {}\or {'}\or {''}\or {'''}\or
   {''''}\or {'''''}\or  \else\@ctrerr\fi\relax}
\newcounter{contador}

\newcommand{\ol}\overline
\newcommand{\ti}\tilde
\newcommand{\wt}\widetilde
\newcommand{\wh}\widehat
\newcommand{\bv}\breve
\newcommand{\dg}\dagger

\newcommand{\be}{\begin{equation}}
\newcommand{\ee}{\end{equation}}
\newcommand{\bl}{\begin{eqnarray}&}
\newcommand{\el}{&\end{eqnarray}}
\newcommand{\bq}{\begin{eqnarray}}
\newcommand{\eq}{\end{eqnarray}}

\renewcommand{\theequation}{\thesection.\arabic{equation}}

\begin{document}

{\hfill
\parbox{50mm}{{\large hep-th/9803247\\
                             CBPF-NF-017/98\\
                             UFES-DF-OP98/2}} \vspace{3mm} }

\begin{center}
{{\LARGE {\bf Exact Scale Invariance of the BF-Yang-Mills Theory\\[3mm]in
Three Dimensions}}}

\vspace{7mm}

{\large Oswaldo M. Del Cima$^{{\rm (a),}}$\footnote{Supported by the 
{\it Fonds zur 
F\"orderung der Wissenschaftlichen Forschung} under the contract 
number P11654-PHY.}, Daniel H.T. Franco$^{{\rm 
(b),}}$\footnote{Supported by the {\it Conselho Nacional de Desenvolvimento 
Cient\'\i fico e Tecnol\'ogico (CNPq)}.},\\[2mm]Jos\'e A. 
Helay\"el-Neto$^{{\rm 
(b),}}$\footnote{Supported in part by the {\it Conselho Nacional de 
Desenvolvimento Cient\'\i fico e Tecnol\'ogico (CNPq)}.}
and Olivier Piguet$^{{\rm (c),3,}}$}\footnote{On leave of 
absence from {\it D\'epartement de Physique Th\'eorique, 
Universit\'e de Gen\`eve, 24 quai E. Ansermet - CH-1211 - Gen\`eve 4 - 
Switzerland}.} 
\vspace{4mm}

$^{{\rm (a)}}$ {\it Technische Universit\"at Wien (TU-Wien),\\
Institut f\"ur Theoretische Physik, \\
Wiedner Hauptstra{\ss}e 8-10 - 
A-1040 - Vienna - Austria.}

$^{{\rm (b)}}$ {\it Centro Brasileiro de Pesquisas F\'\i sicas (CBPF), 
\\Departamento de
Teoria de Campos e Part\'\i culas (DCP),\\Rua Dr. Xavier Sigaud 150 - 
22290-180 - Rio de Janeiro - RJ - Brazil.}

$^{{\rm (c)}}$ {\it Universidade Federal do Esp\'\i rito Santo (UFES), 
\\CCE,
Departamento de F\'\i sica, \\Campus Universit\'ario de Goiabeiras - 
29060-900 - Vit\'oria - ES - Brazil.}
\vspace{4mm} 

{\it E-mails: delcima@tph73.tuwien.ac.at, dfranco@cbpfsu1.cat.cbpf.br, 
helayel@cbpfsu1.cat.cbpf.br, piguet@cce.ufes.br.} 
\end{center}
%%%%%%%%%%%%%%%%%%%%%%%%%%%%%%%%%%%%%%%%%%%%%%%%%%%%%%%%%%
%%%%%%%%%%%%%%%%%%%%%%%%%%%%%%%%%%%%%%%%%%%%%%%%%%%%%%%%%%%%%%%%
\begin{abstract}
The ``extended'' BF-Yang-Mills theory in 3 dimensions, which contains a 
minimally coupled scalar field, is shown to be ultraviolet finite. It obeys 
a trivial Callan-Symanzik equation, with all $% 
\beta$-functions and anomalous dimensions vanishing. The proof is based on
an anomaly-free trace identity valid to all orders of perturbation theory.
\end{abstract}
%%%%%%%%%%%%%%%%%%%%%%%%%%%%%%%%%%%%%%%%%%%%%%%%%%%%%%%%%%%%%%%%

%\newpage

\section{Introduction}

%%%%%%%%%%%%%%%%%%%%%%%%%%%%%%%%%

 Original inspiration for 
topological gauge field theories (TGFT) came from
mathematics. Due to their peculiar properties, they are at the origin of a
large number of interesting results and the object of continuous
investigations.

The topological Yang-Mills theory and Chern-Simons theory are examples of
two distinct classes of TGFT which are sometimes classified as being  of
``Witten type'' and of ``Schwarz type'', respectively (see~\cite{bbr-pr91}
for a general review and references).

Besides the Chern-Simons there exists another TGFT of Schwarz type, namely
the topological BF theory. The latter represents a natural generalization of
the Chern-Simons theory since it can be defined on manifolds of any
dimensions whereas a Chern-Simons action exists only in odd-dimensional
space-times. Moreover, the Lagrangian of the BF theory 
 always contains the
quadratic terms needed for defining a quantum theory, whereas a Chern-Simons
action shows this feature only in three dimensions.

On the other hand, the Yang-Mills gauge theory has  recently been
re-interpreted as a deformation of a pure BF 
theory~\cite{fucito,zeni,mart1,mart2,henn}. 
 In its ``extended'' version~\cite{mart2}, this 
BF-Yang-Mills (BFYM) theory is described by the action 
\begin{equation}
\Sigma _{{\rm BFYM}}=\int d^nx\,\,\left\{ \varepsilon ^{\mu _1\ldots \mu
_n}B_{\mu _1\ldots \mu _{n-2}}^aF_{\mu _{n-1}\mu _n}^a+\left( B_{\mu
_1\ldots \mu _{n-2}}^a+D_{[\mu _1}\eta _{\mu _2\ldots \mu _{n-2}]}^a\right)
^2\right\} \,\,,  \label{action0}
\end{equation}
where $F_{\mu\nu}^a=\partial_{\mu}A_{\nu}^a-
\partial_{\nu}A_{\mu}^a+gf^{abc}A_{\mu}^bA_{\nu}^c$ is
the Yang-Mills field strength, 
$D_\mu\eta_{\mu_1\ldots \mu_{n-3}}^a\equiv
\partial_{\mu}\eta_{\mu_1\ldots \mu_{n-3}}^a + 
gf^{abc}A_{\mu}^b \eta_{\mu_1\ldots \mu_{n-3}}^c$ is the covariant derivative, 
$B_{[\mu_1\ldots \mu _{n-2}]}^a$ an auxiliary
field and  $\eta_{[\mu_1\ldots \mu_{n-3}]}^a$ a pure gauge field.

The action (\ref{action0}) possesses two kinds of symmetries: the standard
gauge symmetry 
\begin{equation}
\delta _\alpha A_\mu ^a=-D_\mu \alpha ^a,\;\;\;\delta _\alpha B_{\mu
_1\ldots \mu _{n-2}}^a=-gf_{\,}^{abc}B_{\mu _1\ldots \mu _{n-2}}^b\alpha
^c,\;\;\;\delta _\alpha \eta _{\mu _1\ldots \mu _{n-3}}=-gf_{\,}^{abc}\eta
_{\mu _1\ldots \mu _{n-3}}^b\alpha ^c\,\,,  \label{sy-gauge}
\end{equation}
and,  thanks to the presence of the field $\eta$,  
the ``topological'' symmetry 
\begin{equation}
\delta _\Lambda A_\mu ^a=0,\;\;\;\delta _\Lambda B_{\mu _1\ldots \mu
_{n-2}}^a=-D_{[\mu _1}\Lambda _{\mu _2\ldots \mu _{n-2}]}^a,\;\;\;\delta
_\Lambda \eta _{\mu _1\ldots \mu _{n-3}}=\Lambda _{\mu _1\ldots \mu
_{n-3}}^a\,\,.  \label{sy-topo}
\end{equation}

It is very easy to verify that if one uses the gauge condition 
$\eta^a_{\mu_1\ldots \mu _{n-3}}=0$ and  the equation of motion 
$\delta \Sigma _{{\rm BFYM}}/\delta B_{\mu _1\ldots \mu _{n-2}}^a=0$, 
one recovers the usual form of the Yang-Mills action.

The quantum equivalence between  the pure Yang-Mills theory 
and the BFYM theory has
been discussed in three dimensions by~\cite{mart2}, and in four dimension by 
\cite{henn,mart3}. In both cases this question has been answered
positively,  thus confirming the interpretation of
the pure YM theory as a
perturbation of a topological theory~\cite{cattaneo}.

The most peculiar property that topological field theories exihibit is their
ultraviolet finiteness~\cite{guada,blasi,magg,lucc}. This property relies
on the existence of a topological vector 
supersymmetry~\cite{bbr-pr91,dgs,silvio}, 
whose origin is manifest in the case one chooses the Landau gauge. 
Actually, such a symmetry turns out to be a general feature of the 
topological theories, and
its explicit realization is extremely simple in the Landau gauge. Moreover,
topological vector supersymmetry 
plays a crucial role in the construction of the
explicit solutions to the BRS cohomology modulo $d$, giving therefore a
systematic classification of all possible anomalies and physically relevant
invariant conterterms~\cite{pigsor}.

 However, there is no such vector supersymmetry in
the present case  since the BFYM theory is not a topological one {\it
stricto sensu}: in particular, the energy-momentum tensor is not 
BRS-exact~\cite{mart2}, a fact which is incompatible with vector
supersymmetry~\cite{silvio}.

 The main purpose of this paper is, starting from the established
renormalizability of the BFYM theory in three dimensions~\cite{mart2},
to give a general proof of its ultraviolet
finiteness, in all orders of perturbation theory. The proof will use the
techniques  developed in~\cite{odhp}. It is based on the 
validity of an anomaly-free trace
identity for the energy-momentum tensor -- i.e. of exact
scale invariance. Actually, we believe that this technique is 
particularly suitable 
in a situation such as our's, where all the power 
of the topological vector  supersymmetry is lost.
%Indeed, this is the case here treated, because the BFYM theory is not a
%topological one -- it can be seen as semi-topological (or topological in the
%field $\eta _{\mu _1\ldots \mu _{n-3}}$ and non-topological in the field $%
%A_\mu ^a$). It can be showed that the energy momentum tensor is not
%BRS-exact~\cite{mart2} -- this is condition which guarantee the topological
%nature of a theory~\cite{bbr-pr91}.

Finally, it should be stressed that the superrenormalizability of the 
 model 
shows to be determinant in the proof of the exact scale invariance of the
BFYM theory in three dimensions.  Indeed, it leaves the
Chern-Simons action -- which has been introduced in order to regularize
the infrared singularity of the theory -- as the only possible invariant
counterterm. Eventually, 
the fact that the integrand of the Chern-Simons action
is not gauge invariant leads to the anomaly-free trace identity.

%%%%%%%%%%%%%%%%%%%%%%%%%%%%%%%%%

\section{The BF-Yang-Mills Theory in Curved Space-Time}

\label{sect2}

\setcounter{equation}{0} %%%%%%%%%%%%%%%%%%%%%%%%%%%%%%%%

Following~\cite{odhp}, we write the BF-Yang-Mills action on a curved
manifold, as long as its topology remains that of flat ${\bf R}^3$, with
asymptotically vanishing curvature. It is the latter two restrictions which
allow us to use the general results of renormalization theory, established
in flat space.

The classical BF-Yang-Mills theory in a curved manifold ${\cal M}$ is
defined by the action 
\begin{equation}
\Sigma _{{\rm BFYM}}=\int d^3x\,\left\{ \varepsilon ^{\mu \nu \rho }B_\mu
^aF_{\nu \rho }^a+e\,\left( B_\mu ^a+D_\mu \eta ^a\right) ^2\right\} \,\,,
\label{action}
\end{equation}
where $e$ denotes the determinant of the dreibein field $e_\mu ^m$. All
fields lie in the adjoint representation of the gauge group, a general
compact Lie group with algebra $\left[ T^a,\,T^b\right] =i\,f^{abc}T^c$.

Following~\cite{mart2}, we add to the action (\ref{action}) a -- parity
breaking -- Chern-Simons term~\cite{deser} 
\begin{equation}
\Sigma _{{\rm CS}}=\int d^3x\,m\,\varepsilon ^{\mu \nu \rho }\left( A_\mu
^a\partial _\nu A_\rho ^a+\frac g3f^{abc}A_\mu ^aA_\nu ^bA_\rho ^c\right)
\,\,,  \label{cs}
\end{equation}
where $m$ is a topological mass. This will safe the theory from IR
divergences. The zero limit, which formally recovers the massless theory, is
argued to be smooth for resummed quantities~\footnote{%
According to~\cite{t'hooft} the observables shoud be independent of the mass 
$m$.}. Moreover, the Chern-Simons term does not change neither the algebraic
structure nor the form of the operators entering the algebraic analysis.

In the Landau gauge, the gauge-fixing term $\Sigma _{{\rm gf}}$ reads 
\begin{equation}
\Sigma _{{\rm gf}}=-\,s\,\displaystyle{\int }d^3x\,\,e\,g^{\mu \nu }\left(
\partial _\mu \bar{c}_aA_\nu ^a+\partial _\mu \bar{\phi}_aB_\nu ^a\right)
\,\,.  \label{g-fix-action}
\end{equation}

The action 
\begin{equation}
\Sigma _{{\rm BFYM}}+\Sigma _{{\rm CS}}+\Sigma _{{\rm gf}}\,\,,
\label{action1}
\end{equation}
is invariant under the nilpotent $s$-operator defined as follows: 
\begin{eqnarray}
sA_\mu ^a &=&-D_\mu c^a\,\,,  \nonumber \\ sc^a &=&\frac 
12\,g\,f^{abc}c^bc^c\,\,,  \nonumber \\ s{\bar{c}}^a &=&b^a\,,  
\nonumber \\ sb^a &=&0\,\,,  \nonumber \\ sB_\mu ^a &=&-D_\mu \phi 
^a\,-g\,f^{abc}B_\mu ^bc^c\,\,,  \label{brs} \\ s\eta ^a &=&\phi 
^a-g\,f^{abc}\eta ^bc^c\,\,,  \nonumber \\ s\phi ^a &=&\,g\,f^{abc}\phi 
^bc^c\,\,,  \nonumber \\ s{\bar{\phi}}^a &=&h^a\,,  
\nonumber \\ sh^a &=&0\,.  \nonumber 
\end{eqnarray}

In order to express the BRS invariance as a Slavnov-Taylor identity we
couple the nonlinear variations of the quantum fields to antifields (or
external sources) $A_a^{*\mu }$, $B_a^{*\mu }$, $c_a^{*}$, $\eta _a^{*}$, $%
\phi _a^{*} $: 
\begin{equation}
\Sigma _{{\rm ext}}=\displaystyle{\int }d^3x\sum_{\Phi =A_\mu ^a,\,B_\mu
^a,\,c^a,\eta ^a,\,\phi ^a}\Phi ^{*}s\Phi \,.  \label{ext-action}
\end{equation}
The total action, 
\begin{equation}
\Sigma =\Sigma _{{\rm BFYM}}+\Sigma _{{\rm CS}}+\Sigma _{{\rm gf}}+\Sigma _{%
{\rm ext}}\,\,,  \label{total action}
\end{equation}
obeys the Slavnov-Taylor identity 
\begin{equation}
{\cal S}(\Sigma )\,\,=\,\,\displaystyle{\int }d^3x\,\displaystyle{\sum_{\Phi
=A_\mu ^a,\,B_\mu ^a,\,c^a,\eta ^a,\,\phi ^a}^{}}{\displaystyle{\frac{\delta
\Sigma }{\delta \Phi ^{*}}}}{\displaystyle{\frac{\delta \Sigma }{\delta \Phi 
}}}\,+b\,\Sigma =0\,,\quad \mbox{with }\ b=\displaystyle{\int }d^3x\,\left(
b^a{{\ \displaystyle{\frac \delta {\delta \bar{c}^a}+}}}h^a{{\ \displaystyle{%
\frac \delta {\delta \bar{\phi}^a}}}}\right) \,.  \label{slavnov}
\end{equation}
For later use, we introduce the linearized Slavnov-Taylor\ operator 
\begin{equation}
{\cal B}_\Sigma \,\,=\,\,\displaystyle{\int }{d}^3x\,\displaystyle{\
\sum_{\Phi =A_\mu ^a,\,B_\mu ^a,\,c^a,\eta ^a,\,\phi ^a}^{}}\left( {%
\displaystyle{\ \frac{\delta \Sigma }{\delta \Phi ^{*}}}}{\displaystyle{%
\frac \delta {\delta \Phi }}}+{\ \displaystyle{\frac{\delta \Sigma }{\delta
\Phi }}}{\displaystyle{\ \frac \delta {\delta \Phi ^{*}}}}\right) \,+b\,.
\label{linear}
\end{equation}
${\cal S}$ and ${\cal B}$ obey the algebraic identity 
\begin{equation}
{\cal B}_{{\cal F}}{\cal B}_{{\cal F}}{\cal F}^{\prime }+\left( {\cal B}_{%
{\cal F}^{\prime }}-b\right) {\cal S}({\cal F})=0\,,  \label{fcondnonab1}
\end{equation}
${\cal F}$ and ${\cal F}^{\prime }$ denoting arbitrary functionals of ghost
number zero. From the latter follows 
\begin{equation}
{\cal B}_{{\cal F}}\,{\cal S}({\cal F})=0\;\;,\;\;\;\forall \;{\cal F}\;\;\;,
\label{nilpot1}
\end{equation}
\begin{equation}
\left( {\cal B}_{{\cal F}}\right) ^2=0\;\;\;{\mbox{if}}\;\;\;{\cal S}({\cal F%
})=0\,.  \label{nilpot3}
\end{equation}
%%%%%%%%%%%%%%%%%%%%%%%%%%%%%%%%%%%%%%%

\subsection{Renormalizability}

%%%%%%%%%%%%%%%%%%%%%%%%%%%%%%%%%%%%%%%
The renormalizability of the theory (absence of anomalies and stability
under the quantum corrections) 
follows\footnote{The authors 
of~\cite{mart2,henn} in fact consider the theory in flat space
only, but the generalization to curved space is straightforward (see for
instance~\cite{odhp} for a similar case).} from the cohomological study 
of~\cite{mart2,henn}. In these papers it is shown indeed that the BRS
cohomology -- in the presence of the antifields -- is empty for what
concerns the anomalies, and corresponds to a possible renormalization of the
physical parameters $g$, $m$ and of the field amplitudes, for what concerns
the counterterms\footnote{%
We shall come back to the question of the counterterms in the next 
Subsection.}. Hence the generating functional 
\[
\Gamma =\Sigma +O(\hbar )
\]
of the vertex functions (one  particle-irreducible Green functions),
considered as a formal power series in $\hbar$, obeys the same
Slavnov-Taylor identity (\ref{slavnov}) as the classical action $\Sigma $: 
\begin{equation}
{\cal S}\left( \Gamma \right) =0\,\,.  \label{qu-slavnov}
\end{equation}

In addition to the Slavnov-Taylor identity, the vertex functional $\Gamma $
obeys the following additional constraints\footnote{See for 
instance~\cite{pigsor}, or in~\cite{odhp} for a proof in a similar 
context.}: 

\begin{itemize}
\item  the Ward identity for the diffeomorphisms 
\begin{equation}
{\cal W}_{{\rm diff}}\Gamma =\displaystyle{\int }d^3x\,\displaystyle{\
\sum_\Phi ^{}}\delta _{{\rm diff}}^{(\varepsilon )}\Phi {\displaystyle{\frac{%
\delta \Gamma }{\delta \Phi }}}=0\,,  \label{diffeo}
\end{equation}
with 
\begin{eqnarray*}
\delta _{{\rm diff}}^{(\varepsilon )}F_\mu =\varepsilon ^\lambda \partial
_\lambda F_\mu +\left( \partial _\mu \varepsilon ^\lambda \right) F_\lambda
\,,\quad F_\mu =\left( A_\mu ^a,\ B_\mu ^a,\,e_\mu ^m\right) \,, \\
\delta _{{\rm diff}}^{(\varepsilon )}\Phi =\varepsilon ^\lambda \partial
_\lambda \Phi \,,\quad \Phi =\left( b^a,\ c^a,\ {\bar{c}}^a,\,h^a,\,\phi
^a,\,\bar{\phi}^a,\,\eta ^a\right) \,, \\
\delta _{{\rm diff}}^{(\varepsilon )}F^{*\mu }=\partial _\lambda \left(
\varepsilon ^\lambda F^{*\mu }\right) -\left( \partial _\lambda
\varepsilon ^\mu \right) F^{*\lambda }\,\,,\quad F^{*\mu }=\left(
A_a^{*\mu },\ B_a^{*\mu }\right) \,, \\
\delta _{{\rm diff}}^{(\varepsilon )}\Phi ^{*}=\partial _\lambda \left(
\varepsilon ^\lambda \Phi ^{*}\right) \,\,\,,\quad \Phi ^{*}=\left(
c^{*a},\,\phi ^{*a},\,\,\eta ^{*a}\right) ;
\end{eqnarray*}

\item  the Ward identity for the local Lorentz transformations 
\begin{equation}
{\cal W}_{{\rm Lorentz}}\Gamma =\displaystyle{\int }d^3x\,\displaystyle{\
\sum_\Phi ^{}}\delta _{{\rm Lorentz}}^{(\lambda )}\Phi {\displaystyle{\frac{%
\delta \Gamma }{\delta \Phi }}}=0\,,  \label{localoren}
\end{equation}
where 
\[
\delta _{{\rm Lorentz}}^{(\lambda )}\Phi =\frac 12\lambda _{mn}\Omega
^{mn}\Phi \,,\quad \Phi =\mbox{any field}\,,
\]
with $\Omega ^{[mn]}$ acting on $\Phi $ as an infinitesimal Lorentz matrix
in the appropriate representation;

\item  the Landau gauge conditions
\[
{\displaystyle{\frac{\delta \Gamma }{\delta b_a}}}=\partial _\mu \left(
eg^{\mu \nu }A_\nu ^a\right) ,
\]
\begin{equation}
\;{\displaystyle{\frac{\delta \Gamma }{\delta h_a}}}=\partial _\mu \left(
eg^{\mu \nu }B_\nu ^a\right) ;  \label{landau}
\end{equation}

\item  the ghost equations of motion 
\[
{\cal G}_{({\rm I})}^a\Gamma ={\displaystyle{\frac{\delta \Gamma }{\delta 
\bar{c}_a}}}+\partial _\mu \left( eg^{\mu \nu }{\displaystyle{\frac{\delta
\Gamma }{\delta A_a^{*\nu }}}}\right) =0\,,
\]
\begin{equation}
{\cal G}_{({\rm II})}^a\Gamma ={\displaystyle{\frac{\delta \Gamma }{\delta 
\bar{\phi}_a}}}+\partial _\mu \left( eg^{\mu \nu }{\displaystyle{\frac{%
\delta \Gamma }{\delta B_a^{*\nu }}}}\right) =0\,;  \label{ghost}
\end{equation}

\item  the antighost equations, (peculiar to the Landau 
gauge~\cite{bl-pig-sor}) 
\[
{\bar{{\cal G}}}_{({\rm I})}^a\Gamma \,\,=\,\,\displaystyle{\int }d^3x\left( 
{\displaystyle{\ \frac \delta {\delta c^a}}}+g\,f^{\,abc}\bar{c}_b{%
\displaystyle{\frac \delta {\delta b^c}+}}g\,f^{\,abc}\bar{\phi}_b{%
\displaystyle{\frac \delta {\delta h^c}}}\right) \Gamma \,\,=\,\,\Delta _{%
{\rm cl}({\rm I})}^a\,,
\]
\begin{equation}
{\bar{{\cal G}}}_{({\rm II})}^a\Gamma \,\,=\,\,\displaystyle{\int }%
d^3x\left( {\displaystyle{\ \frac \delta {\delta \phi ^a}}}+g\,f^{\,abc}\bar{%
\phi}_b{\displaystyle{\frac \delta {\delta b^c}}}\right) \Gamma
\,\,=\,\,\Delta _{{\rm cl}({\rm II})}^a\,,  \label{antighost}
\end{equation}
with 
\[
\Delta _{{\rm cl}({\rm I})}^a=g\displaystyle{\int }d^3x\,f^{\,abc}\left(
A_b^{*\mu }A_{c\mu }+B_b^{*\mu }B_{c\mu }+\eta _b^{*}\eta _c-c_b^{*}c_c-\phi
_b^{*}\phi _c\right) \,,
\]
\[
\Delta _{{\rm cl}({\rm II})}^a=\displaystyle{\int }d^3x\,\left(
gf^{\,abc}\left( B_b^{*\mu }A_{c\mu }-\phi _b^{*}c_c\right) -\eta
^{*a}\right) \,;
\]
(The right-hand sides of (\ref{antighost}) being linear in the quantum
fields will not get renormalized.)

\item  the Ward identities for the rigid gauge  invariances
\begin{equation}
{\cal W}_{{\rm rigid}({\rm I})}^a\Gamma =\displaystyle{\int }d^3x%
\displaystyle{\ \sum_{\Phi \,=\,{\rm all\,fields}}^{}}f^{\,abc}\Phi _b{%
\displaystyle{\frac \delta {\delta \Phi ^c}}}\Gamma =0\,\,,  \label{rigid}
\end{equation}
\[
{\cal W}_{{\rm rigid}({\rm II})}^a\Gamma ={\int }d^3x\,\,\left(
f^{\,abc}\left( A_\mu ^b{{\frac \delta {\delta B_\mu ^c}+\,}}c^b{{\frac
\delta {\delta \phi ^c}+\,}}\bar{\phi}^b{{\frac \delta {\delta \bar{c}^c}+\,}%
}h^b{{\frac \delta {\delta b^c}+B_\mu ^{*b}{{\frac \delta {\delta A_\mu
^{*c}}+\,}}\phi ^{*b}{{\frac \delta {\delta c^{*c}}}}\,}}\right) {{-{{\frac
\delta {\delta \eta ^a}}}}}\right) \Gamma =0\,\,\,,
\]
following from (\ref{slavnov}) and (\ref{antighost}) and from the
``anticommutation relations'' 
\begin{equation}
{\bar{{\cal G}}}_{({\rm A})}^a{\cal S}({\cal F})+{\cal B}_{\cf}\left( {\bar{%
{\cal G}}}_{({\rm A})}^a{\cal F}-\Delta _{{\rm cl}({\rm A})}^a\right) ={\cal %
W}_{{\rm rigid}({\rm A})}^a{\cal F}\ ,\quad {\rm A}={\rm I},{\rm II}\ ,\
\forall \,{\cal F}\ .  \label{antic-gbar-s}
\end{equation}
\end{itemize}

%%%%%%%%%%%%%%%%%%%%%%%%%%%%%%%%%%%%%%%%%%%%%%%%%

\section{Superrenormalizability and Counterterms}

\label{sect3} \setcounter{equation}{0} 

%%%%%%%%%%%%%%%%%%%%%%%%%%%%%%%%%%

The possible invariant counterterms which can be freely added to the action
at each order of perturbation theory are of the trivial or of the nontrivial
type~\cite{pigsor}.  The latter belong to the cohomology 
of the linearized  Slavnov-Taylor operator (\ref{linear}) for the 
integrated insertions and
may be found in~\cite{mart2,henn}. 
It consists of the two gauge invariant terms,
$\int F_{\mu \nu }^aF_a^{\mu \nu }$ and the Chern-Simons action. 
The  trivial counterterms
are all possible variations of field polynomials of dimension 3 at most
and of ghost number $-1$. Ultraviolet dimension, ghost number and Weyl
dimension\footnote{See Section \ref{sect4}.} are displayed in Table 1.

%%%%%%%%%%%%%%%%%%%%%%%%%%%%%%%%%%%%%%%%%%%%%%%%%%%%%%%
\begin{table}[tbh]
\centering
\begin{tabular}{|c||c|c|c|c|c|c|c|c|c|c|c|c|c|c|c|}
\hline
& $A_\mu$ & $B_\mu$ & $\eta$ & $b$ & $h$ & $c$ & ${\overline{c}}$ & $\phi$ & 
${\overline{\phi}}$ & $A^{*\mu }$ & $B^{*\mu}$ & $c^{*}$ & $\eta^{*}$ & $%
\phi^{*}$ &$e_\m^\m$\\ \hline\hline
$d$ & 1/2 & 3/2 & 1/2 & 3/2 & 1/2 & -1/2 & 3/2 & 1/2 & 1/2 & 5/2 & 3/2 & 7/2
& 5/2 & 5/2 &0\\ \hline
$\Phi\Pi$ & 0 & 0 & 0 & 0 & 0 & 1 & -1 & 1 & -1 & -1 & -1 & -2 & -1 & -2 &0\\ 
\hline
$d_{W}$ & -1/2 & 1/2 & 1/2 & 3/2 & 1/2 & -1/2 & 3/2 & 1/2 & 1/2 & 1/2 & -1/2 & 
1/2 & -1 & -1 &-1\\ \hline
\end{tabular}
\caption[t1]{Ultraviolet dimension $d$, ghost number $\Phi \Pi $, and
Weyl dimension $d_W$.}
\label{table2}
\end{table}

%%%%%%%%%%%%%%%%%%%%%%%%%%%%%%%%%%%%%%%%%%%%%%%%%%%%%%

There are however restrictions due to the superrenormalizability of the
theory. The latter stems from the coupling constant-dependent power-counting
formula~\cite{mpr,odhp} 
\begin{equation}
d\left( \gamma \right) =3-\sum\limits_\Phi d_\Phi N_\Phi -\frac 12N_g\,\,,
\label{power}
\end{equation}
for the degree of divergence of a 1-particle irreducible Feynman graph, $%
\gamma $. Here $N_\Phi $ is the number of external lines of $\gamma $
corresponding to the field $\Phi $, $d_\Phi $ is the dimension of $\Phi $ as
given in Table 1, and $N_g$ is the power of the coupling constant $g$ in the
integral corresponding to the diagram $\gamma $.

This formula implies that the dimension of the counterterms is bounded by
three, the dimension 1/2 of the coupling constant $g$ being included in the
computation. But, since they are generated by loop graphs, they are of order
2 in $g$ at least.  Hence, not taking now into account the
dimension of $g$, we  see that their real dimension is bounded by 2.
Imposing also the restrictions due to the gauge 
 conditions (\ref{landau}), ghost equations (\ref{ghost}) and 
antighost equations (\ref{antighost}), as well
as rigid gauge invariance (\ref{rigid}), diffeomorphism invariances 
(\ref{diffeo}) and local Lorentz invariance (\ref{localoren}), 
we easily see that
only the Chern-Simons action survives as a possible counterterm. This  means
that the possible radiative corrections can be reabsorbed through a
redefinition of the topological mass  $m$ only.

%%%%%%%%%%%%%%%%%%%%%%%%%%%%%%%%%%%%%%%%%%%%%%

\section{Quantum Scale Invariance}

\label{sect4} \setcounter{equation}{0} 

%%%%%%%%%%%%%%%%%%%%%

Here, the argument is similar to the one presented in~\cite{odhp}. It relies
on a local form of the Callan-Symanzik equation. This will allow us to
exploit the fact that the integrand of the Chern-Simons term in the action
is not gauge invariant, although its integral is so. Such a local form of
the Callan-Symanzik equation is provided by the ``trace identity''.

The energy-momentum tensor is defined as the tensorial quantum insertion
obtained as the following derivative of the vertex functional with respect
to the dreibein: 
\begin{equation}
{\Theta }_\nu ^{~\mu }\cdot \Gamma =e^{-1}e_\nu ^{~m}~\frac{\delta \Gamma }{%
\delta e_\mu ^{~m}}\,\,.  \label{theta}
\end{equation}
  From the diffeomorphism Ward identity (\ref{diffeo}),  follows the
covariant conservation law: 
\begin{equation}
e\,\nabla _\mu \left[ {\Theta }_\nu ^{~\mu }(x)\cdot \Gamma \right] ={w}_\nu
\left( x\right) \Gamma +\nabla _\mu {w}_\nu ^{~\mu }\left( x\right) \Gamma
\,,  \label{cons-theta}
\end{equation}
where $\nabla _\mu $ is the covariant derivative with respect to the
diffeomorphisms\footnote{For a tensor $T$, such as, e.g., 
$A_\m$ or ${\delta}/{\delta A^{*\mu}}$: 
\[
\nabla_\l T^{\mu\cdots}_{\nu\cdots} = \partial_\l T^{\mu\cdots}_{\nu\cdots}
+ \Gamma_\l{}^\m{}_\rho T^{\rho\cdots}_{\nu\cdots} + \cdots - \Gamma_\l{}^ 
\rho{}_\n T^{\mu\cdots}_{\rho\cdots} - \cdots \,, 
\]
where the $\Gamma_.{}^.{}_.$'s are the Christoffel symbols corresponding to
the connexion $\omega_\m^{mn}$. The covariant derivative of a tensorial
density ${\cal T}$, such as, e.g., $A^{*\mu}$ or $\delta/{\delta A_\m}$, is
related to that of the tensor $e^{-1}{\cal T}\,$ by: 
\[
\nabla_\l {\cal T}^{\mu\cdots}_{\nu\cdots} = e\,\nabla_\l\left( e^{-1} {\cal %
T}^{\mu\cdots}_{\nu\cdots} \right)\,. 
\]
}, the differential operators $w_\lambda \left( x\right) $ and 
$w_\lambda ^{~\mu }\left( x\right) $ acting on $\Gamma $ 
representing contact terms. They are
\begin{equation}
w_\lambda \left( x\right) =\sum\limits_{{\rm all\,fields}}\left( \nabla
_\lambda \Phi \right) \frac \delta {\delta \Phi }\,\,,  \label{a9}
\end{equation}
-- becoming the translation Ward operator in the limit of flat space -- and 
\begin{eqnarray}
w_\lambda ^{~\mu }\left( x\right)  &=&A^{*\mu }\frac \delta {\delta
A^{*\lambda }}+B^{*\mu }\frac \delta {\delta B^{*\lambda }}-A_\lambda \frac
\delta {\delta A_\mu }-B_\lambda \frac \delta {\delta B_\mu }-\delta
_\lambda ^{~\mu }\,\left( c^{*}\frac \delta {\delta c^{*}}+A^{*\nu }\frac
\delta {\delta A^{*\nu }}\right. +  \nonumber \\
&&+\left. B^{*\nu }\frac \delta {\delta B^{*\nu }}+\phi ^{*}\frac \delta
{\delta \phi ^{*}}+\eta ^{*}\frac \delta {\delta \eta ^{*}}\right) \,\,.
\label{a10}
\end{eqnarray}

In the classical theory, the integral of the trace of the tensor $\Theta
_\lambda ^{~\mu }$, 
\begin{equation}
\displaystyle{\int }d^3x\,e\,\Theta _\mu ^{~\mu }=\displaystyle{\int }%
d^3x\,e_\mu ^{~m}\frac{\delta \Sigma }{\delta e_\mu ^{~m}}\equiv N_e\Sigma
\,\,,  \label{int-theta}
\end{equation}
turns out to be an equation of motion, up to soft breakings. This
follows from the identity,  easily checked by inspection of the
classical action: 
\begin{equation}
N_e\Sigma =\left( \sum\limits_{\Phi \,=\,{\rm all\;fields}}d_{{\rm W}}\left(
\Phi \right) N_\Phi +m\partial _m+\frac 12g\partial _g\right) \Sigma \,\,,
\label{a18}
\end{equation}
where $N_\Phi $ is the counting operator and $d_{{\rm W}}\left( \Phi \right) 
$ the Weyl dimensions (see Table 1) of the field $\Phi $. We note that 
 (\ref{a18}) is the Ward identity 
for rigid Weyl symmetry~\cite{iorio} --
broken due to the  dimensionful parameters $m$ and $g$.

The trace $\Theta _\mu ^{~\mu }(x)$ therefore turns out to be vanishing, up
to total derivatives, mass terms and dimensionful coupling, in the classical
approximation,  up to field equations, which means that (\ref{theta}) is
the improved energy-momentum tensor. It is easy to check that, for
the classical theory,  we have in fact: 
\begin{equation}
w(x)\Sigma \equiv \left( e_\mu ^m(x)\frac \delta {\delta e_\mu ^m(x)}-w^{%
{\rm trace}}(x)\right) \Sigma =\Lambda (x)\ ,  \label{ym-cl-tr1}
\end{equation}
or, equivalently: 
\begin{equation}
e\,\Theta _\mu {}^\mu (x)=w^{{\rm trace}}(x)\Sigma +\Lambda (x)\ ,
\label{ym-cl-tr2}
\end{equation}
with 
\begin{equation}
w^{{\rm trace}}\left( x\right) =\sum\limits_{\Phi \,}d_{{\rm W}}\left( \Phi
\right) \Phi \frac \delta {\delta \Phi }\,\,,  \label{a20}
\end{equation}
and 
\begin{eqnarray}
\Lambda  &=&{\displaystyle{\frac m2}}\varepsilon ^{\mu \nu \lambda }A_\mu
^aF_{\nu \lambda }^a-{\displaystyle{\frac g2}}f^{abc}\left( \left( 
A_a^{*\mu }+eg^{\mu \nu }\partial _\nu \bar{c}_a\right) A_\mu ^bc^c+\left( 
B_a^{*\mu }+eg^{\mu \nu }\partial _\nu \bar{\phi}_a\right) B_\mu 
^bc^c-{\frac 12}c^{*\,a}c^bc^c\right. +  \nonumber \\ &&-\left. 
\varepsilon ^{\mu \nu 
\lambda }B_\mu ^aA_\nu ^bA_\lambda ^c-2\,e\,g^{\mu \nu }\left( B_\mu 
^a+D_\mu \eta ^a\right) A_\nu ^b\eta ^c+\phi ^{*a}\phi ^bc^c\right) 
\,+\mbox{total derivative terms}\ , 
\label{lambda}
\end{eqnarray}
$\Lambda $ being invariant under  ${\cal B}_\Sigma$. The latter is the
effect of the breaking of scale invariance due to the dimensionful
parameters. The dimension of $\Lambda $ -- the dimensions of $g$ and $m$ not
being taken into account -- is lower than three: it is a soft breaking.

To promote the trace identity (\ref {ym-cl-tr1}) or (\ref{ym-cl-tr2}) to 
quantum level, we first observe that the following commutation relations 
hold: 
\[
{\cal B}_{{\cal F}}w\left( x\right) {\cal F}-w\left( x\right) {\cal S}({\cal %
F})=0\,\,.
\]
\[
\left[ \frac \delta {\delta \Phi \left( y\right) },w\left( x\right) \right]
=-d_{{\rm W}}\left( \Phi \right) \delta \left( x-y\right) \frac \delta
{\delta \Phi \left( x\right) }\,\,,\;\;\;\Phi =\left( b^a,\,h^a\right) \,\,,
\]
\begin{equation}
\left[ {\bar{{\cal G}}}_{({\rm I})}^a,w(x)\right] =\frac 12\frac \delta
{\delta c^a(x)}\,\,,\;\;\;\left[ {\bar{{\cal G}}}_{({\rm II})}^a,w(x)\right]
=-\left( \frac \delta {\delta \phi ^a\left( x\right) }-gf^{abc}\bar{\phi}%
^b\frac \delta {\delta b^c\left( x\right) }\right) \,\,,  \label{cba}
\end{equation}
\[
\left[ {\cal G}_{({\rm I})}^a\left( y\right) ,w(x)\right] =-\frac 32\delta
\left( x-y\right) {\cal G}_{({\rm I})}^a\left( x\right) +\frac 32\partial
_\mu \delta \left( x-y\right) \left( eg^{\mu \nu }\frac \delta {\delta
A_a^{*\nu }}\right) \left( y\right) \,\,,
\]
\[
\left[ {\cal G}_{({\rm II})}^a\left( y\right) ,w(x)\right] =-\frac 12\delta
\left( x-y\right) {\cal G}_{({\rm II})}^a\left( x\right) +\frac 12\partial
_\mu \delta \left( x-y\right) \left( eg^{\mu \nu }\frac \delta {\delta
B_a^{*\nu }}\right) \left( y\right) \,\,.
\]

Now the relations (\ref{cba}) applied to the vertex functional $\Gamma $
yield, for the insertion $w(x)\Gamma $, the properties 
\[
{\cal B}_\Gamma w\left( x\right) \Gamma =0\,\,,
\]
\[
\frac \delta {\delta b_a(y)}w(x)\Gamma =-\frac 32\partial _\mu \delta
(x-y)\left( e\,g^{\mu \nu }A_\nu ^a\right) (y)\,\,,
\]
\[
\frac \delta {\delta h_a(y)}w(x)\Gamma =-\frac 12\partial _\mu \delta
(x-y)\left( e\,g^{\mu \nu }B_\nu ^a\right) (y)\,\,,
\]
\begin{equation}
{\bar{{\cal G}}}_{({\rm I})}^aw(x)\Gamma =\frac 12\frac{\delta \Gamma }{%
\delta c_a(x)}\,\,,  \label{cba'}
\end{equation}
\[
{\bar{{\cal G}}}_{({\rm II})}^aw(x)\Gamma =-\left( \frac \delta {\delta \phi
^a\left( x\right) }-gf^{abc}\bar{\phi}^b\frac \delta {\delta b^c\left(
x\right) }\right) \Gamma +gf^{abc}\left( B_\mu ^{*b}A^{\mu c}-\phi
^{*b}c^c\right) \left( x\right) 
\]
\[
{\cal G}_{({\rm I})}^a\left( y\right) w(x)\Gamma =\frac 32\partial _\mu \delta
\left( x-y\right) \left( eg^{\mu \nu }\frac{\delta\Gamma}{\delta A_a^{*\nu
}}\right) \left( y\right) 
\]
\[
{\cal G}_{({\rm II})}^a\left( y\right) w(x)\Gamma =\frac 12\partial _\mu \delta
\left( x-y\right) \left( eg^{\mu \nu }\frac{\delta\Gamma}{\delta B_a^{*\nu
}}\right) \left( y\right) 
\]
where we have used the constraints (\ref{landau}), (\ref{ghost}) and (\ref
{antighost}).

The quantum version of (\ref{ym-cl-tr1}) or (\ref{ym-cl-tr2}) will be
written as 
\begin{equation}
w(x)\Gamma =\Lambda (x)\cdot \Gamma +\Delta (x)\cdot \Gamma \,\,,
\label{ym-quant-trace}
\end{equation}
where $\Lambda (x)\cdot \Gamma $ is some quantum extension of the classical
insertion (\ref{lambda}), subjected to the same constraints (\ref{cba'}) as $%
w(x)\Gamma $ (see the Appendix). It follows that the insertion $\Delta \cdot
\Gamma $ defined by (\ref{ym-quant-trace}) obeys the homogeneous constraints 
\[
{\cal B}_\Gamma \left[ \Delta (x)\cdot \Gamma \right] =0\,\,, 
\]
\begin{equation}
\frac \delta {\delta b_a(y)}\left[ \Delta (x)\cdot \Gamma \right]
=0\,\,,\;\;\frac \delta {\delta h_a(y)}\left[ \Delta (x)\cdot \Gamma \right]
=0\,\,,  \label{cond1}
\end{equation}
\[
{\bar{{\cal G}}}_{{\rm (A)}}^a\left[ \Delta (x)\cdot \Gamma \right]
=0\,\,,\,\,\,\,\,{\cal G}_{{\rm (A)}}^a\left[ \Delta (x)\cdot \Gamma \right]
=0\,\,,\;\;\;{\rm A}= {\rm I},\,{\rm II} \,\,, 
\]
beyond the conditions of invariance or covariance under ${\cal W}_{{\rm diff}%
}$, ${\cal W}_{{\rm Lorentz}}$ and ${\cal W}_{{\rm rigid}}$.

By power-counting the insertion $\Delta \cdot \Gamma$ has dimension 3, but
being an effect of the radiative corrections, it  possesses a factor $g^2$ at
least, and thus its effective dimension is at most two due to the
superrenormalizability (see (\ref{power})). It turns out that there is no
insertion obeying all these constraints -- 
 effective power-counting selects the
Chern-Simons Lagrangian density, but the latter is not BRS invariant. Hence $%
\Delta \cdot \Gamma = 0$: there is no radiative correction to the insertion $%
\Lambda \cdot \Gamma$ describing the breaking of scale invariance, and (\ref
{ym-quant-trace}) becomes  
\begin{equation}
e\,{\Theta }_\mu ^{~\mu }(x)\cdot \Gamma =w^{{\rm trace}}(x) \Gamma+\Lambda
(x)\cdot \Gamma \,.  \label{trace-id-tilde}
\end{equation}

This local trace identity leads to a trivial Callan-Symanzik equation  (see
e.g. Section 6 of~\cite{odhp}): 
\begin{equation}
\left( m\partial _m+\frac 12g\partial _g\right)  \Gamma =\int
d^3x~\Lambda (x)\cdot \Gamma \,,
\end{equation}
but now with no radiative effect at all: the $\beta $-functions associated
to the parameters $g$ and $m$ both vanish, scale invariance remaining
affected only by the soft breaking $\Lambda $. There is also no anomalous
dimension  as well.

%%%%%%%%%%%%%%%%%%%%%%%%%%%%%%%%%

\section*{Acknowledgements}
 We thank Manfred Schweda for useful discussions. This 
work has been done in parts at the {\it CBPF-DCP}, at the 
{\it Dept. of Physics} of the
{\it UFMG} and at the the {\it Dept. of Physics} of the {\it UFES}. 
We thank all these three institutions for their financial help for travel 
expenses and for their kind hospitality. One of the authors (O.M.D.C.) 
dedicates this work to his wife, Zilda Cristina, to his daughter, Vittoria, 
and to his son, Enzo, who is coming.

%%%%%%%%%%%%%%%%%%%%%%%%%%%%%%%%%%
%\newpage
%%%%%%%%%%%%%%%%%%%%%%%%%%%%%%%%%%%%%%%%%%%%%%%%%%%%%%%%%%%%%
%%%%%%%%%%%%%%%%%%%%%%%%%%%%%%%%%%%%%%%%%%%%%%%%%%%%%%%%%%%%%
\appendix 
\renewcommand{\theequation}{\Alph{section}.\arabic{equation}} %
\renewcommand{\thesection}{\Alph{section}}

\setcounter{equation}{0} \setcounter{section}{1}

%%%%%%%%%%%%%%%%%%%%%%%%%%%%%%%%%%

\section*{Appendix}

%%%%%%%%%%%%%%%%%%%%%%%%%%%%%%%%%%  

We sketch here a proof that it is possible to construct the quantum
extension $\Lambda \left( x\right) \cdot \Gamma $, appearing in (\ref
{ym-quant-trace}), of the classical insertion $\Lambda \left( x\right) $
given in (\ref{ym-cl-tr2},\ref{lambda}), such that it obeys the same
constraints as $w(x)\Gamma $ in (\ref{cba'}). Let us first introduce a BRS
invariant external field $\rho \left( x\right) $, with dimension and ghost
number zero coupled to $\Lambda \left( x\right) $, introducing the new
classical action 
\begin{equation}
\Sigma ^{\left( \rho \right) }=\Sigma +\int d^3x\,\rho \left( x\right)
\Lambda \left( x\right) \,\,.  \label{app1}
\end{equation}
Defining the ``extended'' antighost operators (see (\ref{antighost})) 
\begin{eqnarray}
{\bar{{\cal G}}}_{({\rm I})}^{\left( \rho \right) a} &=&{\bar{{\cal G}}}_{%
{\rm I}}^a-\frac 12\int d^3x\,\,\rho \frac \delta {\delta c^a}\,\,, 
\nonumber \\
{\bar{{\cal G}}}_{({\rm II})}^{\left( \rho \right) a} &=&{\bar{{\cal G}}}_{(%
{\rm II})}^a+\int d^3x\,\,\rho \left( \frac \delta {\delta \phi ^a}-gf^{abc}%
\bar{\phi}^b\frac \delta {\delta b^c}\right) \,\,,  \label{app2}
\end{eqnarray}
and the new classical insertions 
\begin{eqnarray}
\Delta _{{\rm cl}({\rm I})}^{\left( \rho \right) a} &=&\Delta _{{\rm cl}(%
{\rm I})}^a\,\,,  \nonumber \\
\Delta _{{\rm cl}({\rm II})}^{\left( \rho \right) a} &=&\Delta _{{\rm cl}(%
{\rm II})}^a+g\int d^3x\,\,\rho \,\left( \,f^{abc}\left( B_\mu ^{*b}A^{\mu
c}-\phi ^{*b}c^c\right) -\eta ^{*a}\right) \,\,,  \label{app3}
\end{eqnarray}
we easily see that the fulfillment of the required constraints amounts to
prove, for the new vertex functional 
\begin{equation}
\Gamma ^{\left( \rho \right) }=\Sigma ^{\left( \rho \right) }+O\left( \hbar
\right) \,\,,  \label{app4}
\end{equation}
the Slavnov-Taylor identity 
\begin{equation}
{\cal S}\left( \Gamma ^{\left( \rho \right) }\right) =0\,\,,  \label{app5}
\end{equation}
and the new antighost identities 
\begin{equation}
{\bar{{\cal G}}}_{({\rm I})}^{\left( \rho \right) a}\Gamma ^{\left( \rho
\right) }=\Delta _{{\rm cl}({\rm I})}^{\left( \rho \right) a}+O\left( \rho
^2\right) \,\,,\;\;\;{\bar{{\cal G}}}_{({\rm II})}^{\left( \rho \right)
a}\Gamma ^{\left( \rho \right) }=\Delta _{{\rm cl}({\rm II})}^{\left( \rho
\right) a}+O\left( \rho ^2\right) \,\,,  \label{app6}
\end{equation}
and to impose the extended gauge conditions 
\[
{{\frac{\delta \Gamma ^{\left( \rho \right) }}{\delta b_a}}}=\partial _\mu
\left( eg^{\mu \nu }A_\nu ^a\right) +\frac 32eg^{\mu \nu }\partial _\mu \rho
A_\nu ^a\,\,,
\]
\begin{equation}
{{\frac{\delta \Gamma ^{\left( \rho \right) }}{\delta h_a}}}=\partial _\mu
\left( eg^{\mu \nu }B_\nu ^a\right) +\frac 12eg^{\mu \nu }\partial _\mu \rho
B_\nu ^a\,\,.  \label{app7}
\end{equation}
In (\ref{app6}) and later on, there are terms of order $\rho ^2$ and
more, but we are not interested in them since the constraints we are looking
for are obtained by differentiating (\ref{app5}--\ref{app7}) once with
respect to $\rho $ at $\rho =0$.

The antighost operators (\ref{app2}) forms together with the Slavnov-Taylor
operator (\ref{slavnov}) an algebra 
\[
{\bar{{\cal G}}}_{({\rm I})}^{\left( \rho \right) a}{\cal S}\left( {\cal F}%
\right) +{\cal B}_{{\cal F}}\left( {\bar{{\cal G}}}_{({\rm I})}^{\left( \rho
\right) a} {\cal F} -\Delta _{{\rm cl}({\rm I})}^{\left( \rho \right)
a}\right) ={\cal W}_{{\rm rigid}_{\left( ({\rm I})\right) }}^{\left( \rho
\right) a}{\cal F\,\,}, 
\]
\[
{\bar{{\cal G}}}_{({\rm II})}^{\left( \rho \right) a}{\cal S}\left( {\cal F}%
\right) +{\cal B}_{{\cal F}}\left( {\bar{{\cal G}}}_{({\rm II})}^{\left(
\rho \right) a} {\cal F} -\Delta _{{\rm cl}({\rm II})}^{\left( \rho \right)
a}\right) ={\cal W}_{{\rm rigid}_{\left( ({\rm II})\right) }}^{\left( \rho
\right) a}{\cal F\,\,}, 
\]
\[
\frac \delta {\delta b_a}
\left( {\bar{{\cal G}}}_{({\rm I})}^{\left( \rho
\right) a}{\cal F}-\Delta _{{\rm cl}({\rm I})}^{\left( \rho \right) a}\right) 
-{\bar{%
{\cal G}}}_{({\rm I})}^{\left( \rho \right) a}
\left( \frac {\delta{\cal F}} {\delta
b_a}-\partial _\mu \left( eg^{\mu \nu }A_\nu ^a\right) -\frac 32eg^{\mu \nu
}\partial _\mu \rho A_\nu ^a\right) =0\,\,, 
\]
\[
\frac \delta {\delta h_a}
\left( {\bar{{\cal G}}}_{({\rm II})}^{\left( \rho
\right) a}{\cal F}-\Delta _{{\rm cl}({\rm II})}^{\left( \rho \right) a}\right)
 -{%
\bar{{\cal G}}}_{({\rm II})}^{\left( \rho \right) a}
\left( \frac {\delta{\cal F}}
{\delta h_a}-\partial _\mu \left( eg^{\mu \nu }B_\nu ^a\right) -\frac
12eg^{\mu \nu }\partial _\mu \rho B_\nu ^a\right) =0\,\,, 
\]
\[
{\bar{{\cal G}}}_{({\rm I})}^{\left( \rho \right) a}
\left( {\bar{{\cal G}}}%
_{({\rm I})}^{\left( \rho \right) b}{\cal F}- 
\Delta _{{\rm cl}({\rm I})}^{\left(
\rho \right) b}\right) +{\bar{{\cal G}}}_{({\rm I})}^{\left( \rho \right)
b}\left( {\bar{{\cal G}}}_{({\rm I})}^{\left( \rho \right) a}{\cal F}- 
\Delta _{{\rm %
cl}({\rm I})}^{\left( \rho \right) a}\right) =0\,\,, 
\]
\begin{equation}
{\bar{{\cal G}}}_{({\rm II})}^{\left( \rho \right) a} 
\left( {\bar{{\cal G}}}%
_{ ({\rm II})}^{\left( \rho \right) b}{\cal F}- 
\Delta _{{\rm cl}({\rm II})}^{\left(
\rho \right) b}\right) + {\bar{{\cal G}}}_{({\rm II})}^{\left( \rho \right)
b}\left( {\bar{{\cal G}}}_{({\rm II})}^{\left( \rho \right) a}{\cal F}-
\Delta _{{\rm %
cl}({\rm II})}^{\left( \rho \right) a}\right) =0\,\,,  \label{app9}
\end{equation}
where ${\cal F}$ is an arbitrary functional of ghost number zero, and 
\begin{equation}
{\cal W}_{{\rm rigid}_{\left( ({\rm I})\right) }}^{\left( \rho \right) a}=%
{\cal W}_{{\rm rigid}_{\left( ({\rm I})\right) }}^a\,\,,  \label{app10}
\end{equation}
\[
{\cal W}_{{\rm rigid}_{\left( ({\rm II})\right) }}^{\left( \rho \right) a}={%
\int }d^3x\,\,\left( 1+\rho \right) \left( f^{\,abc}\left( A_\mu ^b{{\frac
\delta {\delta B_\mu ^c}+\,}}c^b{{\frac \delta {\delta \phi ^c}+\,}}\bar{\phi%
}^b{{\frac \delta {\delta \bar{c}^c}+\,}}h^b{{\frac \delta {\delta
b^c}+B_\mu ^{*b}{{\frac \delta {\delta A_\mu ^{*c}}+\,}}\phi ^{*b}{{\frac
\delta {\delta c^{*c}}}}\,}}\right) {{-{{\frac \delta {\delta \eta ^a}}}}}%
\right) \,\,. 
\]

The proof of the Slavnov-Taylor identity (\ref{app5}) follows from the
triviality of the cohomology for the $\rho $-dependent cocycles of ghost
number one. 

The recursive proof of (\ref{app6}) also follows that of the $\rho=0$ theory 
\cite{pigsor}. Assuming (\ref{app6}) to be valid up to the order $n-1$ in $%
\hbar $, we can write 
\begin{equation}
{\bar{{\cal G}}}_{({\rm A})}^{\left( \rho \right) a}\Gamma ^{\left( \rho
\right) }=\Delta_{{\rm cl}({\rm A})}^{\left( \rho \right) a}+\hbar^n
\,\Delta _{({\rm A})}+O\left( \hbar ^{n+1}\right) +O\left( \rho ^2\right)
\,\,,\quad {\rm A} = {\rm I,II}\ ,  \label{app11}
\end{equation}
where $\Delta _{{\rm I}}$  and $\Delta _{{\rm II}}$ are 
integral local field polynomials of dimension 7/2 and 5/2, 
respectively (dimension 1/2 of $g$ included), 
ghost number -1 which, due the algebra (\ref{app9}), are subjected to the
constraints 
\begin{equation}
\frac{\delta \Delta _{({\rm A})}^b}{\delta b_a}=\frac{\delta \Delta _{({\rm A%
})}^b}{\delta h_a}=0\,\,,\;\;\;{\cal B}_{\Gamma ^{\left( \rho \right)
}}\Delta _{({\rm A})}^b=0\,\,,  \label{app12}
\end{equation}
and to the integrability conditions 
\begin{equation}
{\bar{{\cal G}}}_{({\rm A})}^{\left( \rho \right) a} \Delta _{({\rm B})}^b+{%
\bar{{\cal G}}}_{({\rm B})}^{\left( \rho \right) b}\Delta _{({\rm A}%
)}^a=0\,\,.  \label{app13}
\end{equation}
Using the constraints (\ref{app12}), the integrability conditions simplify
to 
\begin{equation}
\int d^3x\,\,\frac \delta {\delta X_{({\rm A})}^a}\Delta _{({\rm B})}^b+\int
d^3x\,\,\frac \delta {\delta X_{({\rm B})}^b}\Delta _{({\rm A})}^a=0\,\,,
\label{app14}
\end{equation}
with 
\begin{equation}
X_{({\rm I})}^a=\left( 1-\frac \rho 2\right) ^{-1}c^a\,\,,\;\;\;X_{({\rm II}%
)}^a=\left( 1+\rho \right) ^{-1}\phi ^a\,\,.  \label{app15}
\end{equation}
The solution of (\ref{app15}) is trivial~\cite{pigsor} 
\begin{equation}
\Delta _{({\rm A})}^a=\int d^3x\,\,\frac \delta {\delta X_{({\rm A})}^a}\hat{%
\Delta}\,\,,  \label{app16}
\end{equation}
with 
\[
\frac{\delta \hat{\Delta}}{\delta b_a}=\frac{\delta \hat{\Delta}}{\delta h_a}%
=0\,\,, 
\]
which shows that the breaking $\Delta _{({\rm A})}^a$ at order $n$ in $\hbar 
$ can be compensated by the counterterm $-\hbar^n \,\hat{\Delta}$.

%%%%%%%%%%%%%%%%%%%%%%%%%%%%%%%%%%%%%%%%%%%%%%%%%%%%%%

\end{document}